# Comment: Performance of Double-Robust Estimators When "Inverse Probability" Weights Are Highly Variable

**James Robins, Mariela Sued, Quanhong Lei-Gomez and Andrea Rotnitzky**

## 1. GENERAL CONSIDERATIONS

We thank the editor Ed George for the opportunity to discuss the paper by Kang and Schaeffer.

The authors' paper provides a review of double-robust (equivalently, double-protected) estimators of (i) the mean $\mu = E(Y)$ of a response $Y$ when $Y$ is missing at random (MAR) (but not completely at random) and of (ii) the average treatment effect in an observational study under the assumption of strong ignorability. In our discussion we will depart from the notation in Kang and Schaeffer (throughout, K&S) and use capital letters to denote random variables and lowercase letter to denote their possible values.

In the missing-data setting (i), one observes $n$ i.i.d. copies of $O = (T, X, TY)$, where $X$ is a vector of always observed covariates and $T$ is the indicator that the response $Y$ is observed. An estimator of $\mu$ is double-robust (throughout, DR) if it remains consistent and asymptotically normal (throughout, CAN) when either (but not necessarily both) a model for the propensity score $\pi(X) \equiv P(T=1|X) = P(T=1|X,Y)$ or a model for the conditional mean $m(X) \equiv E(Y|X) = E(Y|X, T=1)$ is correctly specified, where the equalities follow from the MAR assumption. The authors demonstrate, via simulation, that when a linear logistic model for the propensity score and a linear model for the mean of $Y$ given $X$ are both moderately misspecified, there exists a joint distribution under which the OLS regression estimator $\hat{\mu}_{OLS}$ of $\mu$ outperforms all candidate estimators that depend on a linear logistic maximum likelihood estimate of the propensity score, including all the DR estimators considered by the authors.

Near the end of their Section 1, the authors state that their simulation example "appears to be precisely the type of situation for which the DR estimators of Robins et al. were developed." They then suggest that their simulation results imply that the cited quotation from Bang and Robins (2005) is incorrect or, at the very least, misguided. We disagree with both the authors' statement and suggestion. First, the cited quote neither claims nor implies that when a linear logistic model for the propensity score and a linear model for the mean of $Y$ given $X$ are moderately misspecified, DR estimators always outperform estimators—such as regression, maximum likelihood, or parametric (multiple) imputation estimators—that do not depend on the estimated propensity score. Indeed, Robins and Wang (2000) in their paper "Inference for Imputation Estimators" stated the following:

> If nonresponse is ignorable, a locally semi-parametric efficient estimator is doubly protected; i.e., it is consistent if either a model for nonresponse or a parametric model for the complete data can be correctly specified. On the other hand, consistency of a parametric multiple imputation estimator requires correct specification of a parametric model for the complete data. However, in cases in which the variance of the 'inverse probability' weights is very large, the sampling distribution of a locally semi-parametric efficient (augmented inverse probability of response weighted) estimator can be markedly skew and highly variable, and a parametric imputation estimator may be preferred.

*James Robins is Professor, Department of Epidemiology, Harvard School of Public Health, Boston, Massachusetts 02115, USA e-mail: akaris@hsph.harvard.edu. Mariela Sued is Assistant Professor, Facultad de Ciencias Exactas y Naturales, Universidad de Buenos Aires and CONICET, Argentina. Quanhong Lei-Gomez is a Graduate Student, Department of Biostatistics, Harvard School of Public Health, Boston, Massachusetts 02115, USA. Andrea Rotnitzky is Professor, Department of Economics, Di Tella University, Buenos Aires, Argentina.*

The just-quoted cautionary message of Robins and Wang (2000) is not far from K&S's take-home message. In Section 5 we show that, in the authors' simulation example, the variance of the estimated "inverse probability" weights is very large and the sampling distribution of their candidate DR estimators is skewed and highly variable. It follows that their example is far from the settings Bang and Robins had in mind when recommending the "routine use of DR estimators." Rather, their example falls squarely into the class for which Robins and Wang (2000) cautioned that a parametric imputation estimator may be preferable to DR estimators.

Even prior to Robins and Wang (2000), Robins, Rotnitzky and colleagues had published extensive warnings about, and simulation studies of, the hazards of highly variable "inverse probability" weights (Robins, Rotnitzky and Zhao, 1995, pages 113–115; Scharfstein, Rotnitzky and Robins, 1999, pages 1108–1113), although not specifically for DR estimators. Due to the fact that the problem of highly variable weights was not the focus of their paper and had already been discussed extensively in earlier papers by Robins and colleagues, Bang and Robins (2005) did not repeat Robins and Wang's (2000) cautionary message. In retrospect, had they done so or had the authors been aware of the Robins and Wang article, a misunderstanding could perhaps have been averted.

Whenever the "inverse probability" weights are highly variable, as in K&S's simulation experiment, a small subset of the sample will have extremely large weights relative to the remainder of the sample. In this setting, no estimator of the marginal mean $\mu = E(Y)$ can be guaranteed to perform well. That is why, in such settings, some "argue that inference about the mean $E(Y)$ in the full population should not be attempted," to quote from the authors' discussion. Yet, surprisingly, in the authors' simulation experiment, the regression estimator $\hat{\mu}_{OLS}$ performed very well with a mean squared error (MSE) less than any of their candidate DR estimators, all of which estimated the propensity score by maximum likelihood under a linear logistic model. The explanation is that, whether due to unusual luck or to "cherry-picking," the chosen data-generating distribution was as if optimized to have $\hat{\mu}_{OLS}$ perform well. Indeed, in Section 5, we "deconstruct" the chosen distribution and show that it possesses a number of specific, some rather unusual, features that together served to insure $\hat{\mu}_{OLS}$ would perform well even under K&S's misspecified models.

Now, even were the chosen joint distribution of $(Y, T, X)$ optimized to have $\hat{\mu}_{OLS}$ perform extremely well as an estimator of $E(Y)$ on data $(TY, T, X)$ in which $Y$ is observed only when $T = 1$, such optimization would not guarantee that $\hat{\mu}_{OLS}$ would also perform well on the data $((1-T)Y, T, X)$ in which $Y$ is observed only when $T = 0$. Based on this insight, in Section 5, we repeat K&S's simulation study, except based on data $((1-T)Y, T, X)$ rather than data $(TY, T, X)$, and show that, indeed, $\hat{\mu}_{OLS}$ is now outperformed by all candidate DR estimators in terms of both bias and MSE.

In the analysis of real, as opposed to simulated data, we do not know a priori whether the features of the joint distribution of $(Y, T, X)$ do or do not favor $\hat{\mu}_{OLS}$. Furthermore, with highly variable "inverse probability" weights, we generally cannot learn the answer from the data, owing to poor power. This suggests that, with highly variable weights, a single estimator, whether $\hat{\mu}_{OLS}$ or a single DR estimator of the mean $\mu$ is never adequate even with MAR data; rather an analyst should either "not attempt to make inference about the mean" or else provide a sensitivity analysis (in which models for both the propensity score and the regression of $Y$ on $X$ and estimators of $\mu$ are varied). In Section 6, we sketch a possible approach to sensitivity analysis.

In this discussion we ask the following question: can we find DR estimators that, under the authors' chosen joint distribution for $(Y, T, X)$, both perform almost as well as $\hat{\mu}_{OLS}$ applied to data $(TY, T, X)$ and yet perform better than $\hat{\mu}_{OLS}$ when applied to data $((1-T)Y, T, X)$. In Section 4 we describe the principles we used to search among the set of possible DR estimators and discuss the expected performance of various candidates. We define a general class of DR estimators, which we refer to as "bounded," that contains the DR estimators that perform best in the setting of highly variable "inverse probability" weights. We further subdivide the class of bounded DR estimators into two subclasses—bounded Horvitz–Thompson DR estimators and regression DR estimators. We then describe various scenarios which favor one subclass over the other. We also explain why certain DR estimators perform particularly poorly in settings with highly variable "inverse probability" weights. The performance of our estimators is examined in the simulations reported in Section 5, which both mimics the simulations in S&K and also repeats it but now using data $((1-T)Y, T, X)$.

A major point emphasized by K&S was that, in their simulations, the regression estimator $\hat{\mu}_{OLS}$ outperformed any DR estimator when both their model for the propensity score and for the regression of $Y$ on $X$ (from now on referred to as the "outcome model") were misspecified. However, they restricted attention to linear logistic propensity score models. In Section 3, we show that $\hat{\mu}_{OLS}$ is CAN for $\mu$ when either (but not necessarily both) a linear model for the inverse propensity score, $1/\pi(X) = X^T\alpha$, or a linear model $X^T\beta$ for the conditional mean $E(Y|X)$ is correctly specified. That is, by definition, $\hat{\mu}_{OLS}$ itself is a DR estimator of $\mu$ when the inverse linear model $\pi(X) = 1/(X^T\alpha)$ for the propensity score is substituted for K&S's linear logistic model!

In K&S's simulation experiment, the linear model $X^T\beta$ and the model $\pi(X) = 1/(X^T\alpha)$ are both misspecified. Yet, under their scenario, $\hat{\mu}_{OLS}$ did not "outperform any DR estimator that is CAN when either the regression model $X^T\beta$ or the model $\pi(x) = 1/(x^T\alpha)$ is correctly specified," precisely because $\hat{\mu}_{OLS}$ is one such DR estimator! Of course, the model $\pi(X) = 1/(X^T\alpha)$ would rarely, if ever, be used in practice as the model does not naturally constrain $\pi(X)$ to lie in $[0, 1]$. Nonetheless, understanding that $\hat{\mu}_{OLS}$ is a DR estimator provides important insight into the meaning and theory of double-robustness.

## 2. GENERAL FORM OF DOUBLE-ROBUST ESTIMATORS

The authors note that many different DR estimators exist and give a number of explicit examples. The authors restrict attention to MAR data with missing response. In this setting Robins (2000), Robins (2002), Tan (2006) and van der Laan and Robins (2003) had previously proposed a rather wide variety of DR estimators, in addition to the DR estimators of Robins and colleagues considered by the authors. Moreover, Scharfstein et al. (1999) and van der Laan and Robins (2003) provided general methods for the construction of DR estimators in models with MAR or coarsened at random (CAR) data. Robins and Rotnitzky (2001) described a general approach to the construction of DR estimators (when they exist) in a very large model class that includes all MAR and CAR models as well as certain nonignorable (i.e., non-CAR) missing data models. Recently, van der Laan and Rubin (2006) have developed a general approach called "targeted maximum likelihood" that has overlap with methods in Scharfstein et al. (1999), Robins (2000, 2002) and Bang and Robins (2005) in the setting of missing response data. We will use the general methods of Robins and Rotnitzky (2001) to find a candidate set of DR estimators among which we then search for ones that perform in simulations as well as or better than those discussed by S&K.

Most of the DR estimators of $\mu$ discussed by K&S are of the general form

$$\hat{\mu}_{DR}(\hat{\pi}, \hat{m}) = \mathbb{P}_n\{\hat{m}(X)\} + \mathbb{P}_n\left[\frac{T}{\hat{\pi}(X)}\{Y - \hat{m}(X)\}\right]$$

or

$$\begin{aligned}(1)\quad \hat{\mu}_{B\text{-}DR}(\hat{\pi}, \hat{m}) &= \mathbb{P}_n\{\hat{m}(X)\} \\ &\quad + \frac{\mathbb{P}_n[T/\hat{\pi}(X)\{Y - \hat{m}(X)\}]}{\mathbb{P}_n\{T/(\hat{\pi}(X))\}},\end{aligned}$$

where throughout, $\mathbb{P}_n(A)$ is a shortcut for $n^{-1} \cdot \sum_{i=1}^n A_i$. Robins, Sued, Lei-Gomez and Rotnitzky (2007) show that these estimators are solutions to particular augmented inverse probability weighted (AIPW) estimating equations. The AIPW estimating equations are obtained by applying the general methods of Robins and Rotnitzky (2001) to the simple missing-data model considered by K&S.

Quite generally, to construct $\hat{m}$ and $\hat{\pi}$ we specify (i) a "working" parametric submodel for the propensity score $\pi(X) \equiv \Pr(T = 1|X)$ of the form

$$(2) \qquad \pi(\cdot) \in \{\pi(\cdot; \alpha) : \alpha \in \mathbb{R}^q\},$$

where $\pi(x; \alpha)$ is a known function, for example, $\pi(x; \alpha) = \{1 + \exp(-x^T\alpha)\}^{-1}$ as in S&K and, (ii) a working parametric model for $m(X) \equiv E(Y|X)$ of the form

$$(3) \qquad m(\cdot) \in \{m(\cdot; \beta) : \beta \in \mathbb{R}^p\},$$

where $m(x; \beta)$ is a known function, for example, $m(x; \beta) = x^T\beta$ as in S&K. We then obtain estimators $\hat{\alpha}$ and $\hat{\beta}$ which converge at rate $n^{1/2}$ to some constant vectors $\alpha^*$ and $\beta^*$, which are, respectively, equal to the true value of $\alpha$ and/or $\beta$ when the corresponding working model is correctly specified, and define $\hat{m}(x) \equiv m(x; \hat{\beta})$ and $\hat{\pi}(x) \equiv \pi(x; \hat{\alpha})$. [In fact, under mild additional regularity conditions, the rate $n^{1/2}$ can be relaxed to $n^\zeta$, $\zeta > 1/4$, a fact which is critical when the dimensions $p$ and $q$ of $\beta$ and $\alpha$ are allowed to increase with $n$ as $n^\rho$, $\rho < 1/2$.]

Under regularity conditions, if either (but not necessarily both) (2) or (3) is correct, $\hat{\mu}_{DR}(\hat{\pi}, \hat{m})$ and $\hat{\mu}_{B\text{-}DR}(\hat{\pi}, \hat{m})$ are consistent and asymptotically normal (CAN) estimators of $\mu$.

In the special case in which (a) $\pi(x; \alpha) = \{1 + \exp(-x^T\alpha)\}^{-1}$ and $m(x; \beta) = \Phi(x^T\beta)$ where $\Phi$ is a known canonical inverse link function, and (b) $\hat{\alpha}$ is the MLE of $\alpha$ and $\hat{\beta}$ is the iteratively reweighted least squares estimator of $\beta$ among respondents (throughout denoted as $\hat{\beta}_{REG}$) satisfying

$$(4) \qquad \mathbb{P}_n[TX\{Y - \Phi(X^T\hat{\beta}_{REG})\}] = 0,$$

we shall denote $\hat{m}$ with $\hat{m}_{REG}$ and the resulting DR estimators as $\hat{\mu}_{DR}(\hat{\pi}, \hat{m}_{REG})$ and $\hat{\mu}_{B\text{-}DR}(\hat{\pi}, \hat{m}_{REG})$. When $\Phi$ is the identity, $\hat{\beta}_{REG}$ is thus the OLS estimator of $\beta$. In such case, $\hat{\mu}_{DR}(\hat{\pi}, \hat{m}_{REG})$ is the estimator denoted $\hat{\mu}_{BC\text{-}OLS}$ in S&K and $\hat{\mu}_{B\text{-}DR}(\hat{\pi}, \hat{m}_{REG})$ is the estimator in the display following (8) in S&K.

## 3. $\hat{\mu}_{OLS}$ AS A DR ESTIMATOR UNDER A LINEAR INVERSE PROPENSITY MODEL

As anticipated in Section 1, in this section we will argue that the regression estimator $\hat{\mu}_{OLS}$ is indeed a DR estimator with respect to specific working models. Suppose that we postulate the linear inverse propensity model, that is, in (2) we take $\pi(x; \alpha) = 1/(\alpha^T x)$. It follows from (4) that for any $\alpha \in \Omega$,

$$(5) \qquad \mathbb{P}_n\left[\frac{T}{\pi(X; \alpha)}\{Y - \hat{m}_{REG}(X)\}\right] = 0.$$

Thus, the regression estimator $\hat{\mu}_{OLS}$ is indeed equal to $\hat{\mu}_{DR}(\pi(\cdot; \alpha), \hat{m}_{REG})$ for any $\alpha$ and therefore DR with respect to the linear inverse propensity model $\mathcal{P}_{\text{sub,inv}}$ and the outcome model $\Phi(x^T\beta) = x^T\beta$.

To estimate $\alpha$ in model $\mathcal{P}_{\text{sub,inv}}$ we may use either the estimator $\hat{\alpha}_{\text{inv}}$ or the estimator $\hat{\hat{\alpha}}_{\text{inv}}$ that, respectively, minimize the log-likelihood $\mathbb{P}_n[T\log\{\pi(X; \alpha)\} + \{1-T\}\log\{1-\pi(X; \alpha)\}]$ or squared norm $\|\mathbb{P}_n[\{\frac{T}{\pi(X;\alpha)} - 1\}X]\|^2$ both subject to the constraints $\pi(X_i; \alpha) \geq 0$, $i = 1, \ldots, N$. Under regularity conditions, $\hat{\alpha}_{\text{inv}}$ and $\hat{\hat{\alpha}}_{\text{inv}}$ converge in probability to quantities $\alpha^*_{\text{inv}}$ and $\alpha^{**}_{\text{inv}}$ with the property that when model $\mathcal{P}_{\text{sub,inv}}$ is correctly specified, $\pi(X; \alpha^*_{\text{inv}})$ and $\pi(X; \alpha^{**}_{\text{inv}})$ are equal to the propensity score $P(T = 1|X)$.

## 4. DOUBLE-ROBUST ESTIMATORS WITH DESIRABLE PROPERTIES

### 4.1 Boundedness

We would like to have DR estimators of $\mu = E(Y)$ with the "boundedness" property that, when the sample space of $Y$ is finite, they fall in the parameter space for $\mu$ with probability 1. Neither $\hat{\mu}_{DR}(\hat{\pi}, \hat{m}_{REG})$ nor $\hat{\mu}_{B\text{-}DR}(\hat{\pi}, \hat{m}_{REG})$ has this property. We consider two separate ways to guarantee the "boundedness" property.

First, suppose that we found DR estimators that could be written in the IPW form

$$(6) \qquad \mathbb{P}_n\{YT/\hat{\pi}(X)\}/\mathbb{P}_n\{T/\hat{\pi}(X)\}$$

for some nonnegative $\hat{\pi}(\cdot)$. Then the property would hold for such estimators. Specifically, the quantity in the last display is a convex combination of the observed $Y$-values and thus always lies in the interval $[Y_{\min}, Y_{\max}]$ with endpoints the minimum and maximum observed $Y$-values. But, $[Y_{\min}, Y_{\max}]$ is included in the parameter space for $\mu$ because $\mu$ is the population mean of $Y$.

Note that division by $\mathbb{P}_n\{T/\hat{\pi}(X)\}$ is essential to ensure that (6) is in $[Y_{\min}, Y_{\max}]$. In particular, the Horvitz–Thompson estimator $\hat{\mu}_{HT} = \mathbb{P}_n\{YT/\hat{\pi}(X)\}$ does not satisfy this property. For example, if $Y$ is Bernoulli, (6) lies in $[0, 1]$ but $\hat{\mu}_{HT}$ may lie outside $[0, 1]$. For instance, this will be the case if in the sample there exists a unit with $T = 1$, $\hat{\pi}(X) < 1/n$ and $Y = 1$ since then $\hat{\mu}_{HT}$ will be greater than 1. Indeed, when we carried out 1000 Monte Carlo replications of Kang and Schaeffer's simulation experiment for sample size $n = 1000$ using their misspecified analysis model for $\pi(x)$, we found in one particularly anomalous replication, a simulated unit with $T = 1$ but with the unusually small estimated propensity $\hat{\pi}(X) < 1/17{,}000$. Thus, had we simulated $Y$ from a Bernoulli rather than from a normal distribution, we would have had $\hat{\mu}_{HT} > 17$ for this anomalous replication!

The desire that an estimator falls in the interval $[Y_{\min}, Y_{\max}]$ is in conflict with the desire that it be unbiased, as we now show. Suppose the propensity score function $\pi(x)$ were known. The set of exactly unbiased estimators of $\mu$ (that are invariant to permutations of the index $i$ labeling the units) are contained in the set $\{\mathbb{P}_n[T\{Y - q(X)\}/\pi(X) + q(X)]\}$ as $q(X)$ varies. It follows that no unbiased estimator of $\mu$ exists for $Y$ Bernoulli that is guaranteed to fall in $[Y_{\min}, Y_{\max}]$. Taking $q(X)$ identically zero, we obtain $\widetilde{\mu}_{HT} = \mathbb{P}_n\{YT/\pi(X)\}$. Suppose that in an actual study of 1000 subjects, a rare fluctuation had occurred and there was a subject with $T = 1$ whose propensity score $\pi(X)$ was less than 1/17,000 so $\widetilde{\mu}_{HT} > 17$. We doubt any scientist could be convinced to publish the logically impossible estimate $\widetilde{\mu}_{HT}$ for the mean of a Bernoulli $Y$, with the argument that only then would his estimator of the mean be exactly unbiased over hypothetical repetitions of the study that, of course, neither have occurred nor will occur. Exactly analogous difficulties arise for any other choice of $q(X)$. With highly variable weights, "boundedness" trumps unbiasedness.

Second, the boundedness property also holds for DR estimators that can be written in the regression form

(7) $$\mathbb{P}_n\{\hat{m}(X)\}$$

with $\hat{m}(x) = \Phi(x^T\hat{\beta} + h(x)^T\hat{\gamma})$ for some specified function $h(\cdot)$ and an inverse link function $\Phi$ satisfying $\inf \mathbf{Y} \leq \Phi(u) \leq \sup \mathbf{Y}$ for all $u$, where $\mathbf{Y}$ is the sample space of $Y$. This follows because (i) $\mathbb{P}_n\{\hat{m}(X)\}$ falls in the interval $[\hat{m}_{\min}, \hat{m}_{\max}]$, with $\hat{m}_{\min}$ and $\hat{m}_{\max}$ the minimum and maximum values of $\hat{m}(X)$ among the $n$ sample units and, (ii) the above choice of $\Phi(\cdot)$ guarantees that $[\hat{m}_{\min}, \hat{m}_{\max}]$ is contained in the parameter space for $\mu$.

Neither the estimator $\hat{\mu}_{DR}(\hat{\pi}, \hat{m}_{REG})$ nor the estimator $\hat{\mu}_{B-DR}(\hat{\pi}, \hat{m}_{REG})$ of Section 2 satisfies the "boundedness" property. Note, however, that $\hat{\mu}_{B-DR}(\hat{\pi}, \hat{m})$ satisfies $|\hat{\mu}_{B-DR}(\hat{\pi}, \hat{m})| < \max_{i=1,\ldots,n}|Y_i - \hat{m}(X_i)| + \max_{i=1,\ldots,n}|\hat{m}(X_i)|$. Thus when $Y$ is Bernoulli and $m(X) = \Phi(X^T\beta)$ for $\Phi$ any inverse link with range in $[0,1]$, we have that $|\hat{\mu}_{B-DR}(\hat{\pi}, \hat{m})| < 2$, so it is within a factor of 2 of lying in the parameter space. In contrast, when $Y$ is Bernoulli, $\hat{\mu}_{DR}(\hat{\pi}, \hat{m})$, like $\hat{\mu}_{HT}$, can be extremely large.

In the next sections, we describe general approaches to constructing DR estimators that can be written in the form (6) or the form (7). Thus it is important to determine whether DR estimators that satisfy (6) perform better or worse than those satisfying (7) when models for both $m(\cdot)$ and $\pi(\cdot)$ are wrong. Unfortunately, no general recommendation can be given because the answer will depend on the specific data generating process and models used to estimate $m(\cdot)$ and $\pi(\cdot)$. For example, suppose that as in Kang and Schaeffer's simulation experiment, $Y$ is continuous, $\text{var}(Y|X) = \sigma^2$ does not depend on $X$, $\Phi$ is the identity link and we estimate the model $X^T\beta$ for $m(X) = E(Y|X)$ by OLS. When (i) $\sigma^2/\text{Var}(Y)$ is near zero and (ii) there exist a number of nonrespondent units $j$ (i.e., units $j$ with $T_j = 0$) whose values $x_j$ of $X$ lie far outside the convex hull of the set of values of $X$ for the subsample of respondents, then $y_j$ and $m(x_j)$ will be close to one another but not to $\hat{m}_{REG}(x_j)$ except if, by luck, the model $E(Y|X) = X^T\beta$ is so close to being correct that the fit of the model to the subsample of respondents allows successful linear extrapolation to $X$'s far from those fitted. As we shall see, it is precisely such "luck" that explains the good performance of $\hat{\mu}_{OLS}$ in the authors' simulation experiment. Without such luck, the estimator $\mathbb{P}_n\{\hat{m}_{REG}(X)\}$ may perform poorly compared to $\mathbb{P}_n\{YT/\hat{\pi}(X)\}/\mathbb{P}_n\{T/\hat{\pi}(X)\}$, owing to unsuccessful linear extrapolation. On the other hand, when $\sigma^2/\text{Var}(Y)$ is close to 1 and very few units with $T = 0$ have values of $X$ far outside the convex hull of the set of $X$'s in the respondents subsample, $\mathbb{P}_n\{\hat{m}_{REG}(X)\}$ will generally perform better than $\mathbb{P}_n\{YT/\hat{\pi}(X)\}/\mathbb{P}_n\{T/\hat{\pi}(X)\}$, as the latter may be dominated by the large weights $1/\hat{\pi}(X)$ assigned to respondents who have both very small values of $\hat{\pi}(X)$ and large residuals $Y - m(X)$.

4.1.1 *Regression double-robust estimators.* We refer to DR estimators satisfying (7) as regression DR estimators. They are obtained by replacing $\hat{m}_{REG}$ in $\hat{\mu}_{DR}(\hat{\pi}, \hat{m}_{REG})$ with $\hat{m}(X)$ satisfying

(8) $$\mathbb{P}_n\left[\frac{T}{\hat{\pi}(X)}\{Y - \hat{m}(X)\}\right] = 0.$$

Here we describe three such estimators, though others exist (Robins et al., 2007).

The first one, proposed in Scharfstein et al. (1999) and discussed further in Bang and Robins (2005), is the estimator in K&S's Table 8. To compute this estimator one considers an extended outcome model of the form $\Phi(X^T\beta + \varphi\hat{\pi}(X)^{-1})$ adding the covariate $\hat{\pi}(X)^{-1}$ (i.e., the inverse of the fitted propensity score). One then jointly estimates $(\beta, \varphi)$ with $(\widetilde{\beta}, \widetilde{\varphi})$ satisfying $\mathbb{P}_n[T\{Y - \Phi(X^T\widetilde{\beta} + \widetilde{\varphi}\hat{\pi}(X)^{-1})\}\begin{bmatrix}\hat{\pi}(X)^{-1}\\X\end{bmatrix}] = 0$. The first row of this last equation is precisely (8) with $\hat{m}(X)$ replaced with the fitted regression function $\hat{m}_{EXT\text{-}REG}(X) = \Phi(X^T\widetilde{\beta} + \widetilde{\varphi}\hat{\pi}(X)^{-1})$. Consequently, $\hat{\mu}_{DR}(\hat{\pi}, \hat{m}_{EXT\text{-}REG}) = \mathbb{P}_n\{\hat{m}_{EXT\text{-}REG}(X)\}$. This estimator is CAN provided either the model $\pi(x; \alpha)$ for the propensity score $\pi(x)$ or the model $\Phi(x^T\beta)$ for $E(Y|X = x)$ is correct. In fact, it is CAN even if model $\Phi(x^T\beta)$ is incorrect provided the model $\Phi(X^T\beta + \varphi\{\pi(X;\alpha^*)\}^{-1})$ is correct, where $\alpha^*$ is the probability limit of the estimator $\hat{\alpha}$ of $\alpha$. In particular, as indicated in the previous subsection, if $Y$ is Bernoulli and $\Phi$ is the inverse logit link, $\hat{\mu}_{DR}(\hat{\pi}, \hat{m}_{EXT\text{-}REG})$ is always in $[0,1]$.

Nonetheless, when $Y$ is continuous and $\Phi$ is the identity, $|\hat{\mu}_{DR}(\hat{\pi}, \hat{m}_{EXT\text{-}REG})|$ can be disastrously large when the estimated inverse probability weights $\hat{\pi}(X)^{-1}$ are highly variable. Specifically, when $\hat{\pi}(X)^{-1}$ is highly variable, it could very well happen that in most repeated samples the largest value of $\hat{\pi}^{-1}$ among nonrespondents is manyfold greater than the largest value among the respondent subsample. [E.g., a typical Monte Carlo replication of Kang and Schaeffer under the wrong propensity score model had a largest $\hat{\pi}(X)^{-1}$ of 80 in the respondent subsample but a largest $\hat{\pi}(X)^{-1}$ of 1800 in the nonrespondent subsample.] In such cases, enormously greater extrapolation would be required with model $\Phi\{X^T\beta + \varphi\hat{\pi}(X)^{-1}\}$ than with model $\Phi(X^T\beta)$ to obtain fitted values for $Y$ in the nonrespondent subsample, clearly a problem if the extrapolation model $\Phi\{X^T\beta + \varphi\hat{\pi}(X)^{-1}\}$ is also wrong. This phenomenon explains the disastrous performance of $\hat{\mu}_{DR}(\hat{\pi}, \hat{m}_{EXT\text{-}REG})$ observed in K&S's Table 8 when both the model for the propensity score and the outcome model are wrong.

A second DR estimator with the regression form (7) is immediately obtained by estimating the parameter $\beta$ of the model $E(Y|X) = \Phi(X^T\beta)$ with the weighted least squares estimator $\hat{\beta}_{WLS}$ that uses weights $1/\hat{\pi}(X)$. By definition, the estimator $\hat{\beta}_{WLS}$ satisfies

$$\mathbb{P}_n\left[\frac{T}{\hat{\pi}(X)}\{Y - \Phi(X^T\hat{\beta}_{WLS})\}X\right] = 0$$

and consequently (8) is immediately true for $\hat{m}(X)$ equal to $\hat{m}_{WLS}(X) = \Phi(X^T\hat{\beta}_{WLS})$ when, as we always assume in this discussion, the first component of $X$ is the constant 1. It therefore follows that when model $\Phi(X^T\beta)$ has an intercept, $\hat{\mu}_{DR}(\hat{\pi}, \hat{m}_{WLS}) = \mathbb{P}_n(\hat{m}_{WLS})$. The estimator $\hat{\mu}_{DR}(\hat{\pi}, \hat{m}_{WLS})$ is called $\hat{\mu}_{WLS}$ in K&S.

With highly variable $\hat{\pi}(X)^{-1}$ and incorrect models for both $\pi(X)$ and $E(Y|X)$, we would expect $\hat{\mu}_{DR}(\hat{\pi}, \hat{m}_{WLS})$ to outperform $\hat{\mu}_{DR}(\hat{\pi}, \hat{m}_{EXT\text{-}REG})$ because it does not have the severe extrapolation problem of $\hat{\mu}_{DR}(\hat{\pi}, \hat{m}_{EXT\text{-}REG})$. This expectation is dramatically borne out in K&S's simulations.

Some years ago, Marshall Joffe pointed out to us that $\hat{\mu}_{DR}(\hat{\pi}, \hat{m}_{WLS})$ was double-robust and asked us if it had advantages compared to $\hat{\mu}_{DR}(\hat{\pi}, \hat{m}_{EXT\text{-}REG})$. At the time we had not realized that $\hat{\mu}_{DR}(\hat{\pi}, \hat{m}_{EXT\text{-}REG})$ would perform so very poorly in settings with highly variable $\hat{\pi}(X)^{-1}$, so we told him that it probably offered no particular advantage. Based on our bad advice, Joffe never published a paper on $\hat{\mu}_{DR}(\hat{\pi}, \hat{m}_{WLS})$ as a DR estimator. To our knowledge, Kang and Schaeffer are the first to do so. We note that Kang and Schaeffer do not consider $\hat{\mu}_{DR}(\hat{\pi}, \hat{m}_{WLS})$ to be an AIPW DR estimator. However, the above derivation shows otherwise.

Even $\hat{\mu}_{DR}(\hat{\pi}, \hat{m}_{WLS})$ may not perform well in some instances. For example, if $\text{Var}(Y|X) = \sigma^2$ is constant, $\sigma^2/\text{Var}(Y)$ is near 1 and a number of nonrespondents have $X$ lying far outside the convex hull of the respondents' $X$ values, then $\hat{\mu}_{DR}(\hat{\pi}, \hat{m}_{WLS})$ may perform poorly. This is because the subjects who have the greatest $\hat{\pi}(X)^{-1}$ in the respondents' subsample will have enormous leverage which can force their residual to be nearly zero, which is a problem particularly when $\sigma^2/\text{Var}(Y)$ is near 1 and the model $\Phi(X^T\beta)$ is misspecified, as then extrapolation to the $X$'s far from the convex hull will be poor.

The third DR regression type estimator is an extension of the estimator $\hat{\mu}_{IPW\text{-}NR}$ in Kang and Schaeffer. To compute this estimator we extend the regression model $\Phi(X^T\beta)$ by adding the covariate $\hat{\pi}(X)$ (rather than its inverse) to obtain $\Phi\{X^T\beta + \varphi\hat{\pi}(X)\}$ and then jointly estimate $(\beta, \varphi)$ with the estimator $(\widetilde{\beta}, \widetilde{\varphi})$ satisfying

$$\mathbb{P}_n\left[\frac{T}{\hat{\pi}(X)}\{Y - \Phi(X^T\widetilde{\beta} + \widetilde{\varphi}\hat{\pi}(X))\}\begin{bmatrix}\hat{\pi}(X) \\ X\end{bmatrix}\right] = 0.$$

Because we have assumed the vector $X$ has one component equal to the constant 1, (8) is satisfied with $\hat{m}(X)$ equal to $\hat{m}_{DR\text{-}IPW\text{-}NR}(X) = \Phi\{X^T\widetilde{\beta} + \widetilde{\varphi}\hat{\pi}(X)\}$. Thus, $\hat{\mu}_{DR}(\hat{\pi}, \hat{m}_{DR\text{-}IPW\text{-}NR}) = \mathbb{P}_n(\hat{m}_{DR\text{-}IPW\text{-}NR})$. Furthermore, since by construction, $\mathbb{P}_n[T\{Y - \hat{m}_{DR\text{-}IPW\text{-}NR}(X)\}] = 0$, then $\hat{\mu}_{DR}(\hat{\pi}, \hat{m}_{DR\text{-}IPW\text{-}NR})$ is also equal to $\mathbb{P}_n\{TY + (1-T)\hat{m}_{DR\text{-}IPW\text{-}NR}(X)\}$. Because $\hat{\pi}(X)$ is bounded between 0 and 1, adding the covariate $\hat{\pi}(X)$ to model $\Phi(X^T\beta)$, in contrast to adding $\hat{\pi}(X)^{-1}$, does not induce model extrapolation problems like the ones discussed above for $\hat{\mu}_{DR}(\hat{\pi}, \hat{m}_{EXT\text{-}REG})$. We speculate that $\hat{\mu}_{DR}(\hat{\pi}, \hat{m}_{DR\text{-}IPW\text{-}NR})$ will behave much better than $\hat{\mu}_{DR}(\hat{\pi}, \hat{m}_{EXT\text{-}REG})$ and possibly similarly to $\hat{\mu}_{DR}(\hat{\pi}, \hat{m}_{WLS})$ when $\hat{\pi}(X)^{-1}$ has high variance. Indeed, we have observed this behavior in the simulation study of Section 5; however, due to space limitations, results for $\hat{\mu}_{DR}(\hat{\pi}, \hat{m}_{EXT\text{-}REG})$ were not reported as they were qualitatively similar to those reported in K&S.

Finally, the estimator $\hat{\mu}_{DR}(\hat{\pi}, \hat{m}_{EXT\text{-}REG})$ with $\Phi$ the identity link is also an example of a DR targeted maximum likelihood estimator of the marginal mean $\mu$ in the sense of van der Laan and Rubin (2006). We thus conclude from the above discussion that with highly variable $\hat{\pi}(X)^{-1}$ and incorrect parametric models for $\pi(X)$ and $E(Y|X)$, certain targeted maximum likelihood estimators can perform much worse than the ad hoc estimator $\hat{\mu}_{DR}(\hat{\pi}, \hat{m}_{WLS})$.

4.1.2 *Bounded Horvitz–Thompson double-robust estimators.* We refer to DR estimators satisfying (6) as bounded Horvitz–Thompson DR estimators. They are obtained by replacing $\hat{\pi}$ in $\hat{\mu}_{B\text{-}DR}(\hat{\pi}, \hat{m}_{REG})$ with $\hat{\pi}_{EXT}$ satisfying

$$(9) \quad \mathbb{P}_n\left[(\hat{m}_{REG}(X) - \hat{\mu}_{REG})\left(\frac{T}{\hat{\pi}_{EXT}(X)} - 1\right)\right] = 0$$

where $\hat{\mu}_{REG} = \mathbb{P}_n[\hat{m}_{REG}(X)]$. We can obtain such a $\hat{\pi}_{EXT}(\cdot)$ by considering the extended logistic model $\pi_{EXT}(X) = \text{expit}\{\alpha^T X + \varphi h(X)\}$ with $h(X)$ a user-supplied function and estimating $\varphi$ with $\widetilde{\varphi}_{PROP\text{-}GREED}$ solving

$$\mathbb{P}_n\left[\left\{\frac{T}{\text{expit}(\hat{\alpha}^T X + \varphi h(X))} - 1\right\}\right.$$
$$\left. \cdot \{\hat{m}_{REG}(X) - \hat{\mu}_{OLS}\}\right] = 0$$

where $\hat{\alpha}$ is the MLE of $\alpha$ in the model $\pi(X) = \text{expit}(\alpha^T X)$. Then $\hat{\pi}_{EXT}(X) = \text{expit}\{\hat{\alpha}^T X + \widetilde{\varphi}_{PROP\text{-}GREED} h(X)\}$ satisfies (9) and consequently $\hat{\mu}_{B\text{-}DR}(\hat{\pi}_{EXT}, \hat{m}_{REG})$ is of the form (6). A default choice for $h(X)$ would be $\hat{m}_{REG}(X) - \hat{\mu}_{OLS}$.

Interestingly, the OLS estimator $\hat{\mu}_{OLS}$ can be viewed not only as a DR estimator, as seen in Section 3, but also as a bounded Horvitz–Thompson DR estimator! Specifically, suppose that $\hat{\hat{\alpha}}_{\text{inv}}$ is used to estimate $\alpha$ in the propensity model $\pi(X; \alpha)$ as in Section 3, except that without imposing the constraints, so that indeed, $\hat{\hat{\alpha}}_{\text{inv}}$ solves $0 = \mathbb{P}_n[\{\frac{T}{\pi(X;\alpha)} - 1\}X]$. Then

$$\mathbb{P}_n\left[\left\{\frac{T}{\pi(X; \hat{\hat{\alpha}}_{\text{inv}})} - 1\right\}(\hat{m}_{REG}(X) - \hat{\mu}_{OLS})\right] = 0$$

and $\hat{\mu}_{OLS}$ is equal to the inverse probability weighted estimator

$$\mathbb{P}_n\left\{\frac{T}{\pi(X; \hat{\hat{\alpha}}_{\text{inv}})} Y\right\} \Big/ \mathbb{P}_n\left\{\frac{T}{\pi(X; \hat{\hat{\alpha}}_{\text{inv}})}\right\}.$$

Robins (2001) and van der Laan and Rubin (2006) describe particular bounded Horvitz–Thompson DR estimators $\hat{\mu}^{(\infty)}$ that are obtained by iterating to convergence a sequence $\hat{\mu}^{(j)}$ of estimators that are not themselves bounded Horvitz–Thompson estimators. However, the Robins (2001) estimator performed poorly in our simulations (results not shown) and no simulation study of the van der Laan and Rubin (2006) estimator has been published to our knowledge with highly variable inverse probability weights. In fact, van der Laan and Rubin (2006) describe an estimator $\hat{\mu}^{(\infty)}$ that is simultaneously a bounded Horvitz–Thompson and a regression DR estimator that is obtained by iterating to convergence a sequence $\hat{\mu}^{(j)}$ of estimators without this dual property. Again we do not know of a simulation study showing that the estimator generally performs well in practice with highly variable inverse probability weights.

## 5. SIMULATION STUDIES

To investigate the nature of the surprisingly good performance of the regression estimator $\hat{\mu}_{OLS}$ in the simulation study of K&S and to evaluate the performance of the additional estimators described in Section 4, we replicated the simulation study of K&S. Table 1 reports the Monte Carlo bias, variance and mean squared error for twelve different estimators of $\mu = 210$, sample sizes $n = 200$ and $1000$ and the four possible combinations of model specifications for the propensity score and the conditional mean of the response given the covariates (the latter referred to throughout as the outcome model).

The estimators reported in rows 1 and 9, rows 3 and 11, rows 4, 12, 17 and 22 and rows 5, 13, 18 and 23 are, respectively, the estimators $\hat{\mu}_{OLS}, \hat{\mu}_{IPW\text{-}POP}$, $\hat{\mu}_{BC\text{-}OLS}$ and $\hat{\mu}_{WLS}$ investigated by K&S. Throughout we use the notational conventions of Sections 2–4, and thus we rename $\hat{\mu}_{BC\text{-}OLS}$ and $\hat{\mu}_{WLS}$ with $\hat{\mu}_{DR}(\hat{\pi}, \hat{m}_{REG})$ and $\hat{\mu}_{DR}(\hat{\pi}, \hat{m}_{WLS})$, respectively. The estimator $\hat{\mu}_{HT}$ is the Horvitz–Thompson type estimator $\mathbb{P}_n\{TY/\hat{\pi}(X)\}$. All remaining estimators are DR estimators of $\mu$ and are defined in Sections 2 and 4.

When both working models are correct, theory indicates that all DR estimators are CAN, asymptotically equivalent, and efficient in the class of estimators that are CAN even if the outcome model is incorrect. We were not surprised then to find that all DR estimators reported in rows 4–8 of Table 1 performed identically and were more efficient than the inefficient IPW estimators $\hat{\mu}_{IPW\text{-}POP}$ and $\hat{\mu}_{HT}$. However, the near-identical behavior of the regression estimator $\hat{\mu}_{OLS}$ caught our attention. The estimator $\hat{\mu}_{OLS}$ is the maximum likelihood estimator of $\mu$, and hence efficient, in a semiparametric model that assumes a parametric form for the conditional mean of $Y$ given the covariates. Thus, we would have expected it to have smaller variance than that of the DR estimators of $\mu$, because when both the propensity score model and the regression model are correct, the latter attains the semiparametric variance bound in the less restrictive (nonparametric) model that does not impose restrictions on the conditional mean of $Y$. A closer examination of the data generating process used by K&S explains this unusual behavior. Under S&K data generating process $Y_i = 210 + 13.7 Z_i^* + \varepsilon_i$, where $Z_i^* = 2Z_{1i} + \sum_{j=2}^{4} Z_{ji}$, with $Z_{ji}, j = 1, \ldots, 4$, and $\varepsilon_i$ mutually independent $N(0, 1)$ random variables. But under this process $Z_i^*$, and hence $Z_i = (Z_{1i}, Z_{2i}, Z_{3i}, Z_{4i})$, is an essentially perfect predictor of $Y_i$: the residual variance $\text{var}(Y_i | Z_i^*)$ is equal to $\text{var}(Y_i)/195$. This striking feature of K&S data generating process is illustrated in Figure 1. The figure shows a scatterplot of $Y$ versus the predicted values from the fit of the correct outcome model to the respondents in a random sample of $n = 200$ units. Dark dots correspond to data points of respondents. White dots correspond to the simulated missing outcomes $Y_i$ of the nonrespondents plotted against the predicted values $Z_i' \hat{\beta}$. The white dots follow nearly perfectly a straight line through the origin and with slope 1: the predicted values are essentially perfect predictors of the missing outcomes! When the outcome and propensity score models are correctly specified, the asymptotic variance of the DR estimator is equal to $\text{var}(Y) + \text{var}[\pi(Z)\{1 - \pi(Z)\}^{-1} \text{var}(Y|Z)]$. When $Z$ is a perfect predictor of $Y$, this variance reduces to $\text{var}(Y)$, the variance of the standardized distribution of $\hat{\mu}_{FULL}$, the sample mean of $Y$ of respondents and nonrespondents. This is not surprising because it is well known that, when the outcome model is correctly specified, a DR estimator asymptotically extracts all the information available in $Z$ to predict $Y$. Since the regression estimator $\hat{\mu}_{OLS}$ cannot be more efficient than $\hat{\mu}_{FULL}$, we conclude that $\hat{\mu}_{OLS}$ and the DR estimators should have nearly identical variance when $Z$ is an almost perfect predictor of $Y$ and indeed this variance should be also almost the same as that of the infeasible estimator $\hat{\mu}_{FULL}$. In our study we had simulated the outcomes of the nonrespondents. Thus, we were indeed able to compute $\hat{\mu}_{FULL}$ and its Monte Carlo variance. As expected, the Monte Carlo variance of $\hat{\mu}_{FULL}$ was essentially the same as that of $\hat{\mu}_{OLS}$ for both sample sizes.

TABLE 1
*Results for simulation study as in K&S*

| Row | Estimator | Sample size 200 | | | Sample size 1000 | | |
|---|---|---|---|---|---|---|---|
| | | Bias | Var | MSE | Bias | Var | MSE |
| Both models right | | | | | | | |
| 1 | $\hat{\mu}_{OLS}$ | 0.13 | 5.97 | 5.98 | −0.03 | 1.41 | 1.41 |
| 2 | $\hat{\mu}_{HT}$ | −0.08 | 148.92 | 148.92 | 0.17 | 26.46 | 26.49 |
| 3 | $\hat{\mu}_{IPW\text{-}POP}$ | −0.06 | 14.12 | 14.13 | −0.03 | 3.43 | 3.43 |
| 4 | $\hat{\mu}_{DR}(\hat{\pi}, \hat{m}_{REG})$ | 0.13 | 5.96 | 5.98 | −0.03 | 1.41 | 1.41 |
| 5 | $\hat{\mu}_{DR}(\hat{\pi}, \hat{m}_{WLS})$ | 0.13 | 5.97 | 5.98 | −0.03 | 1.41 | 1.41 |
| 6 | $\hat{\mu}_{DR}(\hat{\pi}, \hat{m}_{DR\text{-}IPW\text{-}NR})$ | 0.13 | 5.97 | 5.98 | −0.03 | 1.41 | 1.41 |
| 7 | $\hat{\mu}_{B\text{-}DR}(\hat{\pi}, \hat{m}_{REG})$ | 0.13 | 5.96 | 5.98 | −0.03 | 1.41 | 1.41 |
| 8 | $\hat{\mu}_{B\text{-}DR}(\hat{\pi}_{EXT}, \hat{m}_{REG})$ | 0.13 | 5.97 | 5.98 | −0.03 | 1.41 | 1.41 |
| Both models wrong | | | | | | | |
| 9 | $\hat{\mu}_{OLS}$ | −0.39 | 10.91 | 11.06 | −0.83 | 2.19 | 2.88 |
| 10 | $\hat{\mu}_{HT}$ | 16.87 | 4110.86 | 4395.39 | 38.97 | 39933 | 41452 |
| 11 | $\hat{\mu}_{IPW\text{-}POP}$ | 1.67 | 73.39 | 76.17 | 4.81 | 108.86 | 131.95 |
| 12 | $\hat{\mu}_{DR}(\hat{\pi}, \hat{m}_{REG})$ | −4.90 | 145.93 | 169.91 | −13.91 | 6853.68 | 7047.12 |
| 13 | $\hat{\mu}_{DR}(\hat{\pi}, \hat{m}_{WLS})$ | −2.01 | 10.70 | 14.74 | −2.98 | 2.20 | 11.08 |
| 14 | $\hat{\mu}_{DR}(\hat{\pi}, \hat{m}_{DR\text{-}IPW\text{-}NR})$ | −1.76 | 11.82 | 14.90 | −2.49 | 1.81 | 8.02 |
| 15 | $\hat{\mu}_{B\text{-}DR}(\hat{\pi}, \hat{m}_{REG})$ | −3.82 | 40.07 | 54.65 | −8.03 | 128.61 | 193.13 |
| 16 | $\hat{\mu}_{B\text{-}DR}(\hat{\pi}_{EXT}, \hat{m}_{REG})$ | −2.25 | 11.77 | 16.82 | −3.33 | 3.44 | 14.54 |
| $\pi$-model right, outcome model wrong | | | | | | | |
| 17 | $\hat{\mu}_{DR}(\hat{\pi}, \hat{m}_{REG})$ | 0.55 | 11.82 | 12.12 | 0.07 | 2.81 | 2.82 |
| 18 | $\hat{\mu}_{DR}(\hat{\pi}, \hat{m}_{WLS})$ | 0.65 | 8.82 | 9.24 | 0.16 | 1.90 | 1.93 |
| 19 | $\hat{\mu}_{DR}(\hat{\pi}, \hat{m}_{DR\text{-}IPW\text{-}NR})$ | 0.06 | 7.39 | 7.40 | −0.10 | 1.58 | 1.59 |
| 20 | $\hat{\mu}_{B\text{-}DR}(\hat{\pi}, \hat{m}_{REG})$ | 0.56 | 11.51 | 11.83 | 0.08 | 2.79 | 2.80 |
| 21 | $\hat{\mu}_{B\text{-}DR}(\hat{\pi}_{EXT}, \hat{m}_{REG})$ | 0.53 | 9.41 | 9.69 | 0.11 | 2.08 | 2.09 |
| $\pi$-model wrong, outcome model right | | | | | | | |
| 22 | $\hat{\mu}_{DR}(\hat{\pi}, \hat{m}_{REG})$ | 0.14 | 5.95 | 5.97 | −0.03 | 1.77 | 1.77 |
| 23 | $\hat{\mu}_{DR}(\hat{\pi}, \hat{m}_{WLS})$ | 0.13 | 5.97 | 5.98 | −0.03 | 1.41 | 1.41 |
| 24 | $\hat{\mu}_{DR}(\hat{\pi}, \hat{m}_{DR\text{-}IPW\text{-}NR})$ | 0.13 | 5.97 | 5.98 | −0.03 | 1.41 | 1.41 |
| 25 | $\hat{\mu}_{B\text{-}DR}(\hat{\pi}, \hat{m}_{REG})$ | 0.13 | 5.96 | 5.97 | −0.02 | 1.43 | 1.43 |
| 26 | $\hat{\mu}_{B\text{-}DR}(\hat{\pi}_{EXT}, \hat{m}_{REG})$ | 0.13 | 5.96 | 5.98 | −0.02 | 1.42 | 1.42 |

Theory also indicates that the IPW estimators $\hat{\mu}_{IPW\text{-}POP}$ and $\hat{\mu}_{HT}$ of rows 2 and 3 should be CAN. However, in our simulations, these estimators were nearly unbiased but their sampling distribution was skewed to the right and had very large variance. Figure 2 shows smooth density estimators for these sampling distributions for sample sizes $n = 200$ and $n = 1000$. The skewness and large variance of the IPW estimators were caused by few samples which had respondents with large values of $Y$ and very large weights $1/\hat{\pi}$. Specifically, in most samples, the true $\pi$ values of the respondents were not too small, and consequently the weights $1/\hat{\pi}$ not too large, precisely because by the very definition of $\pi$, having a respondent with a small $\pi$ is a rare event. In the data generating process of K&S, $\pi(Z)$ is negatively correlated with $Y$; the correlation is roughly equal to $-0.6$. Thus, in most samples, the $1/\hat{\pi}$-weighted mean of the $Y$ values of the respondents tended to be smaller than $\mu$. However, in a few samples, some anomalous respondent had a small value of $\pi$. In the computation of $\hat{\mu}_{HT}$, this anomalous respondent carried an unusually large weight $1/\hat{\pi}$ and because his $Y$ value tended to be larger than the mean $\mu$, the estimator $\hat{\mu}_{HT}$ in those rare samples tended to be substantially larger than $\mu$. The skewness lessens as the sample size increases because with large sample sizes, the number of samples which have respondents with small values of $\pi$ also increases. The skewness is also substantially less severe for $\hat{\mu}_{IPW\text{-}POP}$ compared to that of $\hat{\mu}_{HT}$ also as expected since, as discussed in Section 4.1, in any given sample, $|\hat{\mu}_{IPW\text{-}POP}|$ is bounded by the largest observed $|Y|$ value.

Although the Monte Carlo sampling distribution of the IPW estimators gives a rough idea of the shape of the true sampling distribution of these estimators, neither the Monte Carlo bias nor the Monte Carlo variance should be trusted. One thousand replications are not enough to capture the tail behavior of highly skewed sampling distributions, and as such cannot produce reliable Monte Carlo estimates of bias, much less of variance.

Turn now to the case in which the propensity score model is correct but the outcome model is incorrect. Theory indicates that the DR estimators of rows 17 to 21 of Table 1 should be CAN. However, in our simulations nearly all the DR estimators were slightly biased upward. Nevertheless, all DR estimators performed as well as or better, in terms of MSE, than the OLS estimator of row 9.

Consider now the case in which the propensity score model is incorrect but the outcome model is correct. Once again, the almost identical performance of all DR estimators in rows 22–26 of Table 1 with that of the OLS estimator of row 1 is no surprise after recalling that $Z$ is a perfect predictor of $Y$. Specifically, the fact that $Z_i^*$ is a nearly perfect predictor of $Y_i$ implies that $\hat{m}(Z_i)$ is almost identical to the outcome $Y_i$ regardless of whether unit $i$ is a respondent or a nonrespondent and regardless of whether $\hat{m}(Z_i)$ was fit by ordinary least squares or by weighted least squares. Thus, the average of $\hat{m}(Z_i)$ is essentially the same as $\hat{\mu}_{FULL}$ and the sum of $T_i \hat{\pi}_i^{-1}(Y_i - \hat{m}(Z_i))$ is almost zero regardless of the model under which $\hat{\pi}$ was computed. Consequently, all DR estimators must be nearly the same as the infeasible full data sample mean $\hat{\mu}_{FULL}$.

Finally, turn to the case in which both propensity score and the outcome models are wrong. The performance of the IPW estimators is disastrous as well as that of the DR estimator in row 12 and, to a lesser extent, that of the estimator in row 15. Figure 3 shows smooth density estimators of the sampling distribution of these four estimators when the sample size is 1000. The estimators $\hat{\mu}_{HT}$ and $\hat{\mu}_{IPW\text{-}POP}$ have distributions heavily skewed to the right while the estimators $\hat{\mu}_{DR}(\hat{\pi}, \hat{m}_{REG})$ and $\hat{\mu}_{B\text{-}DR}(\hat{\pi}, \hat{m}_{REG})$ have distributions heavily skewed to the left. The skewness is far more dramatic for the estimators $\hat{\mu}_{HT}$ and $\hat{\mu}_{DR}(\hat{\pi}, \hat{m}_{REG})$ than for their counterparts $\hat{\mu}_{IPW\text{-}POP}$ and $\hat{\mu}_{B\text{-}DR}(\hat{\pi}, \hat{m}_{REG})$, reflecting the fact that $\hat{\mu}_{IPW\text{-}POP}$ and $\hat{\mu}_{B\text{-}DR}(\hat{\pi}, \hat{m}_{REG})$ are bounded in the sense described in Section 4.1 while $\hat{\mu}_{HT}$ and $\hat{\mu}_{DR}(\hat{\pi}, \hat{m}_{REG})$ are unbounded. [Indeed, to avoid distortions, in constructing the density plots of $\hat{\mu}_{HT}$ and $\hat{\mu}_{DR}(\hat{\pi}, \hat{m}_{REG})$ we have omitted the extreme values of 5873 and $-2213$, respectively, from one simulation replication.] Rows 12 and 14 of Table 1 report that the Monte Carlo bias and variance indeed are even larger for $n = 1000$ than for $n = 200$. The extreme distribution skewness and the increase in bias and variance with sample size are explained as follows. As noted earlier, even when the $\pi$'s are estimated from a correct model, the distribution of $\hat{\mu}_{HT}$ and $\hat{\mu}_{IPW\text{-}POP}$ will tend to be skewed to the right when $1/\pi$ is positively correlated with $Y$ because of the presence of a few unusual samples with anomalous respondents with large $Y$ values and small $\pi$ values. Now, because of the nature of the wrong analytic propensity score model used in the simulations, the estimated $\hat{\pi}$'s corresponding to the anomalous units in the unusual samples were many times smaller than the true $\pi$'s. As a consequence the, usually large, values of $Y$ of the anomalous units essentially determined the values of $\hat{\mu}_{HT}$ and $\hat{\mu}_{IPW\text{-}POP}$ in the unusual samples and consequently, exacerbated even more the skewness of the Monte Carlo sampling distribution of the IPW estimators. The larger bias and variance when $n = 1000$ than when $n = 200$ were due to two replications with sample size 1000 in which the values of the estimators were extreme [specifically, $\hat{\mu}_{HT} = 1475$ and 2884, and $\hat{\mu}_{DR}(\hat{\pi}, \hat{m}_{REG}) = -2213$ and $-175$]. These outlying values were caused by one anomalous nonrespondent in each sample with large values of $Y$ (the second largest $Y$ values in one sample and the largest in the other). For these units, the $1/\pi$ values were 38.7 and 50.9 but $1/\hat{\pi}$ were 17,068 and 399, respectively. When the two samples with these anomalous units were removed, the variance of the estimators $\hat{\mu}_{HT}$ and $\hat{\mu}_{DR}(\hat{\pi}, \hat{m}_{REG})$ decreased to 6729 and 890, respectively. The paradoxical increase in Monte Carlo variance with sample size is but another proof that the Monte Carlo variance in simulations with 1000 replications is not a reliable estimator of the true variance for estimators with highly skewed distributions. The different directionality of the skewness of the IPW and DR estimators is explained as follows. In the computation of $\hat{\mu}_{DR}(\hat{\pi}, \hat{m}_{REG})$ and $\hat{\mu}_{B\text{-}DR}(\hat{\pi}, \hat{m}_{REG})$ we inverse probability weight the values of $(\hat{m}_{REG} - Y)$. Consequently since, as indicated below, under K&S's wrong analytic outcome model, $\hat{m}_{REG}$ was reasonably bounded; thus, in the few unusual samples, the anomalous units with small $\pi$'s had large and negative values of $(\hat{m}_{REG} - Y)$ and produced extremely small values of the DR estimators.

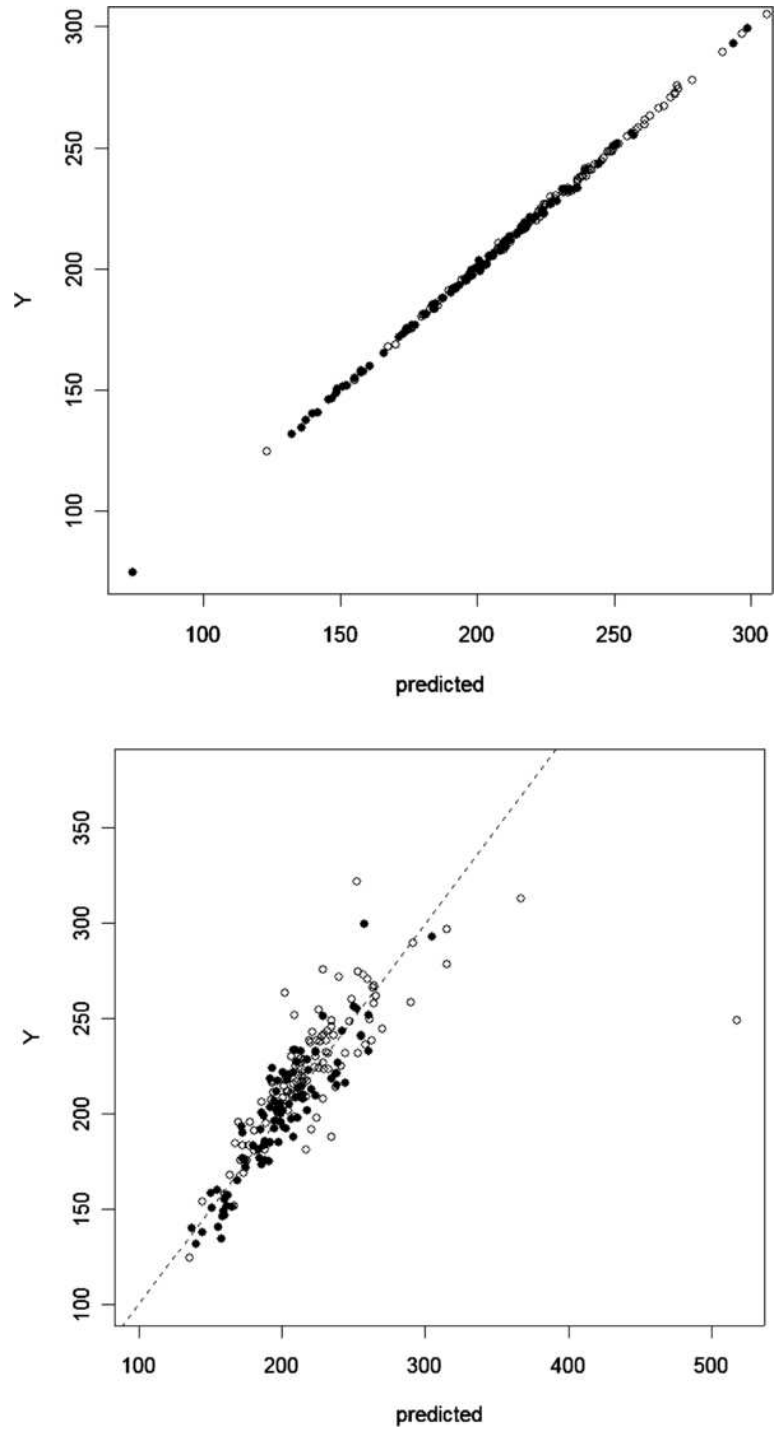

FIG. 1. *K&S simulation experiment. Outcomes vs predicted values. Sample size 200. Top: correct y model. Bottom: wrong y model. Dashed line is the line $Y = X$. Dark dots: respondents. White dots: nonrespondents.*

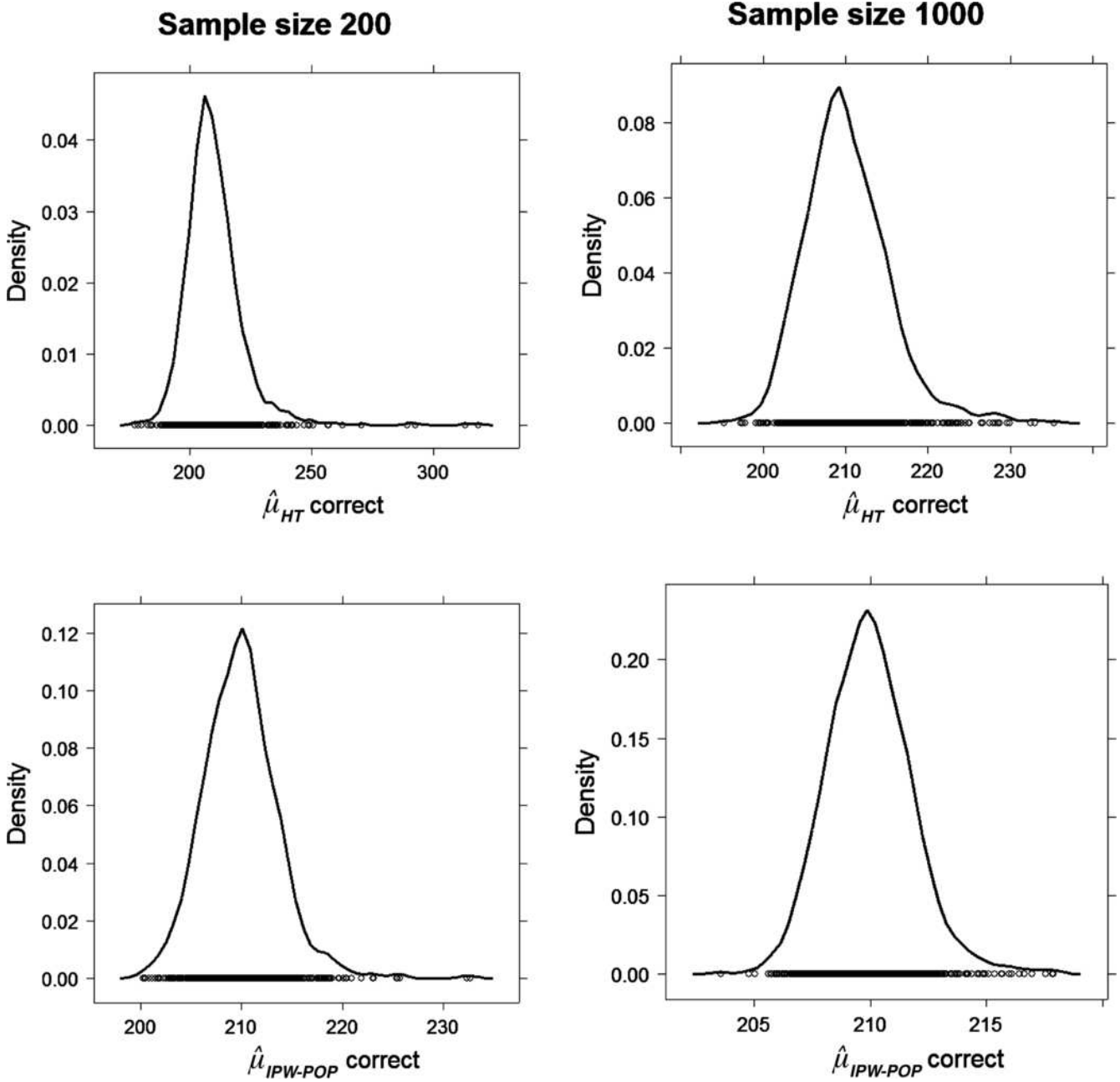

Fig. 2. *Distributions of $\hat{\mu}_{HT}$ and $\hat{\mu}_{IPW\text{-}POP}$ under correct propensity score models.*

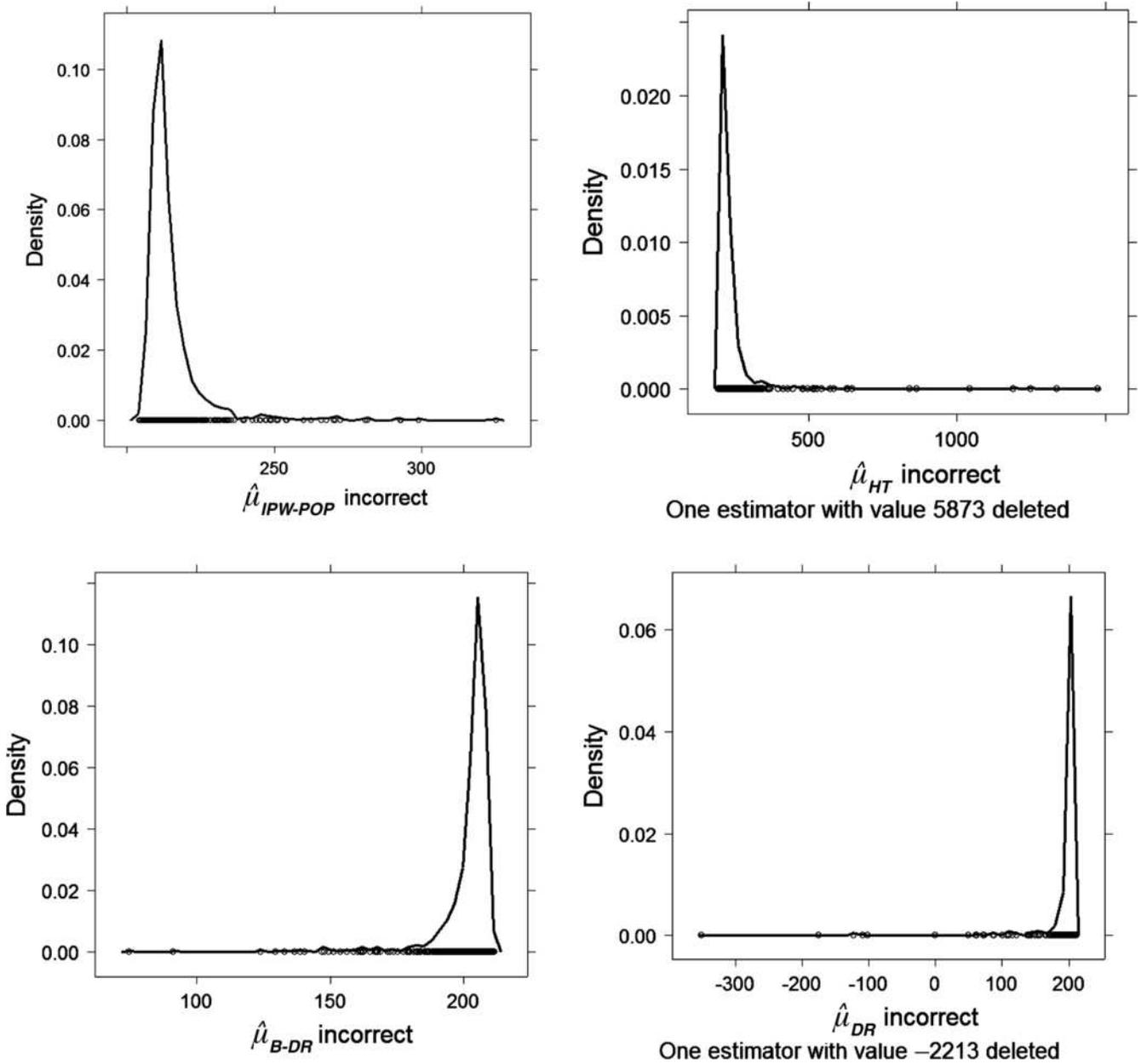

FIG. 3. Distributions of $\hat{\mu}_{HT}, \hat{\mu}_{IPW\text{-}POP}, \hat{\mu}_{DR}(\hat{\pi}, \hat{m}_{REG})$ and $\hat{\mu}_{B\text{-}DR}(\hat{\pi}, \hat{m}_{REG})$ under incorrect propensity score and outcome models.

The performance of the remaining DR estimators in rows 13, 14 and 16 is heterogeneous. Some, though still biased, have bias and variance orders of magnitude smaller than the variance of the estimators $\hat{\mu}_{DR}(\hat{\pi}, \hat{m}_{REG})$ and $\hat{\mu}_{B\text{-}DR}(\hat{\pi}, \hat{m}_{REG})$.

In a second simulation experiment described below, the relative performance of the DR estimators was somewhat different than in this simulation study and, as we explain later, better than that of the regression estimator $\hat{\mu}_{OLS}$. This attests to the obvious fact that when the propensity score and outcome models are both incorrect we cannot expect to find a single clear winner. The relative performance of the estimators will very much depend on the data generating process and the nature of the model misspecifications.

To understand why the regression estimator $\hat{\mu}_{OLS}$ performed so remarkably well when both models were wrong, we first note that because the outcome model was a linear regression model with an intercept fitted by ordinary least squares in the respondent subsample, the sum of the predicted values $X'\hat{\beta}$ and the sum of $Y$ in the respondent subsample are the same. Thus, $\hat{\mu}_{OLS} = (n_{\text{obs}}/n)\overline{Y}_{\text{obs}} + (n_{\text{miss}}/n) \overline{(X'\hat{\beta})}_{\text{miss}}$, where $\overline{(X'\hat{\beta})}_{\text{miss}}$ is the average of the predicted values for the missing outcomes. The bias of $\hat{\mu}_{OLS}$ therefore depends on the bias of $\overline{(X'\hat{\beta})}_{\text{miss}}$ as an estimator of the mean of $Y$ in the nonrespondent subpopulation. If, due either to good luck or "cherry picking," the prediction function $x'\hat{\beta}$ from a misspecified regression model $x'\beta$ successfully extrapolates to the covariates of nonrespondents, even when these are far from the convex hull of covariates in the respondent subsample, $\overline{(X'\hat{\beta})}_{\text{miss}} - \overline{Y}_{\text{miss}}$ will be roughly centered around 0, and consequently $\hat{\mu}_{OLS}$ will be a nearly unbiased estimator of the mean of $Y$. We now show this phenomenon explains the excellent performance of $\hat{\mu}_{OLS}$ in Kang & Schaeffer's simulation. In Figure 4 we plotted the outcomes $Y$ versus the predicted values $X'\hat{\beta}$ in the previously mentioned unusual sample of size 1000 with both the propensity and outcome models misspecified, where $\hat{\mu}_{DR}(\hat{\pi}, \hat{m}_{REG}), \hat{\mu}_{B\text{-}DR}(\hat{\pi}, \hat{m}_{REG})$ and the IPW estimators did disastrously due to the presence of one anomalous unit with extremely small $\hat{\pi}$. The dark dots correspond to the observed data values of the respondents. White dots correspond to the actual simulated missing outcomes $Y$ of the nonrespondents plotted against the predicted values $X'\hat{\beta}$. We can see that the predicted values of the nonrespondents are reasonably centered around the straight line even for those points with predicted values far from the predicted values of the respondents. In this sample, $\hat{\mu}_{OLS}$ was 205.78, a far more reasonable value than those obtained for the IPW and just-mentioned DR estimators.

To demonstrate that $\hat{\mu}_{OLS}$ can have a substantially worse performance than the DR estimators, we conducted a second simulation experiment. This second experiment, like our first, redid K&S's simulation by generating the data $(Y, T, X)$ from K&S's chosen distributions. However, in our second experiment we analyzed the data $((1-T)Y, T, X)$ in which $Y$ is observed only when $T = 0$, rather than the data $(TY, T, X)$ that was analyzed by us in our first experiment and by K&S in their paper. [To do so, since the data $((1-T)Y, T, X)$ can be recoded as $((1-T)Y, 1-T, X)$, we simply recompute each of the estimators reported in Table 1 except now we everywhere replace $\hat{\pi}$ and $T$ by $1 - \hat{\pi}$ and $1 - T$.] The results are displayed in Table 2. We observe that, with both models wrong, the bias and MSE of $\hat{\mu}_{OLS}$ now exceed those of any DR estimator!

As in our first experiment, due to the extreme variability in the estimated "inverse probability" weights, the DR estimators appear to have considerable finite sample bias, especially at the smaller sample size of 200, when the propensity model is correct but the outcome model is wrong. In fact, this bias is larger than it was in the first simulation experiment, which was to be expected as the variability in the estimated "inverse probability" weights was greater in the second than the first experiment (data not shown).

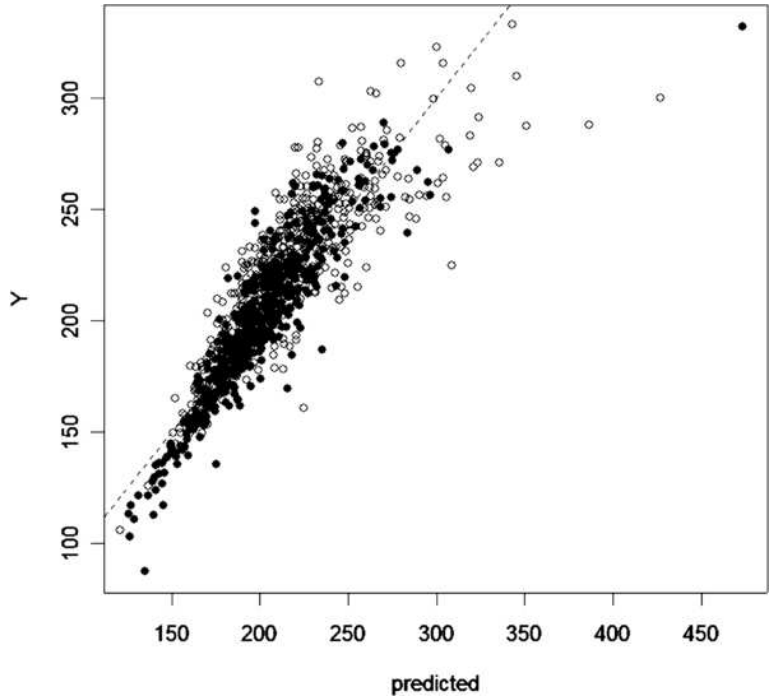

FIG. 4. *Y vs predicted values in one sample of size 1000 generated under K&S experiment. Dashed line is the line $Y = X$. Dark dots: respondents. White dots: nonrespondents.*

## 6. SENSITIVITY ANALYSIS

Consider again the missing-data setting with the mean $\mu$ of $Y$ as the parameter of interest. When the covariate vector $X$ is high dimensional, one cannot be certain, owing to lack of power, that a chosen model for the propensity score is nearly correct, even if it passes standard goodness-of-fit tests. Therefore a large number of models for the propensity score with different subsets of the covariates, different orders of interactions and different dimensions of the parameter vector should be fit to the data. Similarly, many different outcome models should be fit. This raises the question: once fit, how should these many candidate models be used in the estimation of the mean of $Y$?

One approach is to use modern techniques of model selection to choose a single propensity and outcome model. Specifically, there has been a recent outpouring of work on model selection in regression. This work has shown that one can use cross-validation and/or penalization to empirically choose, from among a large number of candidates, a model whose predictive risk for the response variable in the regression is close to that of the best candidate model. In fact, van der Laan (2005) has proposed that $k$-fold cross-validation should be routinely employed to select the model for the propensity score and for the outcome regression that are to be used in the construction of a DR estimator.

An alternative approach which we are currently studying is the following. Suppose one has fit $J_p$ propensity score models and $J_o$ outcome models. For a favorite DR estimator $\hat{\mu}$, define $\hat{\mu}_{ij}$ as the DR estimator that uses the fitted values from the $i$th propensity model and the $j$th outcome model. Now, if the $i$th propensity model is correct, all $J_o$ estimators in the set $\mathcal{E}_{p,i} \equiv \{\hat{\mu}_{ij}; j = 1, \ldots, J_o\}$ will be CAN estimators of $\mu$. Thus, an $\alpha$-level test of the homogeneity hypothesis $H_{pi}: E^A(\hat{\mu}_{i1}) = E^A(\hat{\mu}_{ij})$ for all $j \in \{2, \ldots, J_o\}$ [where $E^A(\cdot)$ stands for large sample mean, i.e., the probability limit of $\cdot$] is also an $\alpha$-level goodness-of-fit test for the propensity model that is directly relevant to its use in a DR estimator of $\mu$. Similarly if the $j$th outcome model is correct, all $J_p$ estimators in the set $\mathcal{E}_{o,j} \equiv \{\hat{\mu}_{ij}; i = 1, \ldots, J_p\}$ will be CAN for $\mu$ and a test of the homogeneity hypothesis $H_{oj}: E^A(\hat{\mu}_{1j}) = E^A(\hat{\mu}_{ij})$ for all $i \in \{2, \ldots, J_p\}$ is a test of fit for the outcome model. This suggests that one could choose as a final estimator of $\mu$ the DR estimator $\hat{\mu}_{i^*j^*}$, where $i^*$ is the $i$ for which the test of the hypothesis $H_{pi}$ gave the largest $p$-value and $j^*$ is the $j$ for which the test of the hypothesis $H_{oj}$ gave the largest $p$-value. However, this method of selecting $i^*$ and $j^*$ is nonoptimal for two reasons. First, it could easily select a misspecified propensity model $i$ for which the power of the test of the hypothesis $H_{pi}$ is particularly poor and similarly for the outcome regression. This remark implies that some measure of the spread of the elements of $\mathcal{E}_{p,i}$ and $\mathcal{E}_{o,j}$ should also contribute to the selection of $i^*$ and $j^*$. Second, the method does not exploit the fact that if $i^*$ and $j^*$ are correct, then $E^A(\hat{\mu}_{ij^*}) = E^A(\hat{\mu}_{i^*j})$ for all $i$ and $j$, suggesting that an optimal method should select $i^*$ and $j^*$ jointly. Alternative approaches for selecting $i^*$ and $j^*$ will be reported elsewhere. In any case, the very fact that input to the selection algorithm requires the matrix $\hat{\mu}_{ij}$ provides an informal sensitivity analysis; we directly observe the sensitivity of our DR estimator to the choice of propensity and outcome regression model.

TABLE 2
*Results for simulation study as in K&S but with the roles of T and 1 − T reversed*

| Row | Estimator | Sample size 200 | | | Sample size 1000 | | |
|---|---|---|---|---|---|---|---|
| | | Bias | Var | MSE | Bias | Var | MSE |
| Both models right | | | | | | | |
| 1 | $\hat{\mu}_{OLS}$ | 0.12 | 5.96 | 5.98 | −0.03 | 1.41 | 1.41 |
| 2 | $\hat{\mu}_{HT}$ | −0.46 | 49.14 | 49.36 | −0.24 | 8.45 | 8.51 |
| 3 | $\hat{\mu}_{IPW\text{-}POP}$ | 0.45 | 14.76 | 14.96 | 0.05 | 3.11 | 3.12 |
| 4 | $\hat{\mu}_{DR}(\hat{\pi}, \hat{m}_{REG})$ | 0.12 | 5.96 | 5.98 | −0.02 | 1.41 | 1.41 |
| 5 | $\hat{\mu}_{DR}(\hat{\pi}, \hat{m}_{WLS})$ | 0.12 | 5.96 | 5.97 | −0.02 | 1.41 | 1.41 |
| 6 | $\hat{\mu}_{DR}(\hat{\pi}, \hat{m}_{DR\text{-}IPW\text{-}NR})$ | 0.12 | 5.96 | 5.97 | −0.02 | 1.40 | 1.41 |
| 7 | $\hat{\mu}_{B\text{-}DR}(\hat{\pi}, \hat{m}_{REG})$ | 0.12 | 5.96 | 5.97 | −0.02 | 1.41 | 1.41 |
| 8 | $\hat{\mu}_{B\text{-}DR}(\hat{\pi}_{EXT}, \hat{m}_{REG})$ | 0.12 | 5.96 | 5.97 | −0.02 | 1.40 | 1.41 |
| Both models wrong | | | | | | | |
| 9 | $\hat{\mu}_{OLS}$ | 4.97 | 7.97 | 32.68 | 4.97 | 1.91 | 26.62 |
| 10 | $\hat{\mu}_{HT}$ | 0.55 | 40.27 | 40.57 | 0.39 | 6.27 | 6.43 |
| 11 | $\hat{\mu}_{IPW\text{-}POP}$ | 3.92 | 9.67 | 25.03 | 3.68 | 2.22 | 15.79 |
| 12 | $\hat{\mu}_{DR}(\hat{\pi}, \hat{m}_{REG})$ | 3.33 | 8.79 | 19.90 | 3.07 | 2.12 | 11.53 |
| 13 | $\hat{\mu}_{DR}(\hat{\pi}, \hat{m}_{WLS})$ | 3.17 | 8.21 | 18.24 | 2.81 | 1.97 | 9.84 |
| 14 | $\hat{\mu}_{DR}(\hat{\pi}, \hat{m}_{DR\text{-}IPW\text{-}NR})$ | 3.11 | 8.21 | 17.90 | 2.64 | 1.97 | 8.94 |
| 15 | $\hat{\mu}_{B\text{-}DR}(\hat{\pi}, \hat{m}_{REG})$ | 3.32 | 8.70 | 19.69 | 3.04 | 2.10 | 11.34 |
| 16 | $\hat{\mu}_{B\text{-}DR}(\hat{\pi}_{EXT}, \hat{m}_{REG})$ | 3.30 | 8.68 | 19.55 | 3.01 | 2.10 | 11.16 |
| $\pi$-model right, outcome model wrong | | | | | | | |
| 17 | $\hat{\mu}_{DR}(\hat{\pi}, \hat{m}_{REG})$ | 0.71 | 12.60 | 13.11 | 0.14 | 2.96 | 2.98 |
| 18 | $\hat{\mu}_{DR}(\hat{\pi}, \hat{m}_{WLS})$ | 0.99 | 8.04 | 9.02 | 0.23 | 1.92 | 1.97 |
| 19 | $\hat{\mu}_{DR}(\hat{\pi}, \hat{m}_{DR\text{-}IPW\text{-}NR})$ | 0.71 | 7.26 | 7.76 | 0.18 | 1.72 | 1.75 |
| 20 | $\hat{\mu}_{B\text{-}DR}(\hat{\pi}, \hat{m}_{REG})$ | 0.75 | 11.21 | 11.76 | 0.14 | 2.76 | 2.78 |
| 21 | $\hat{\mu}_{B\text{-}DR}(\hat{\pi}_{EXT}, \hat{m}_{REG})$ | 0.86 | 10.38 | 11.12 | 0.18 | 2.71 | 2.74 |
| $\pi$-model wrong, outcome model right | | | | | | | |
| 22 | $\hat{\mu}_{DR}(\hat{\pi}, \hat{m}_{REG})$ | 0.12 | 5.96 | 5.97 | −0.02 | 1.40 | 1.41 |
| 23 | $\hat{\mu}_{DR}(\hat{\pi}, \hat{m}_{WLS})$ | 0.12 | 5.96 | 5.97 | −0.02 | 1.40 | 1.41 |
| 24 | $\hat{\mu}_{DR}(\hat{\pi}, \hat{m}_{DR\text{-}IPW\text{-}NR})$ | 0.12 | 5.96 | 5.97 | −0.02 | 1.40 | 1.41 |
| 25 | $\hat{\mu}_{B\text{-}DR}(\hat{\pi}, \hat{m}_{REG})$ | 0.12 | 5.96 | 5.97 | −0.02 | 1.40 | 1.41 |
| 26 | $\hat{\mu}_{B\text{-}DR}(\hat{\pi}_{EXT}, \hat{m}_{REG})$ | 0.12 | 5.96 | 5.97 | −0.02 | 1.40 | 1.41 |

The approach just described could also be combined with the model selection approach. Specifically, one first uses cross-validation to choose not one but rather $J_p$ and $J_o$ propensity and outcome models (the ones with the $J_p$ and $J_o$ lowest cross-validated risk estimates) out of a much larger number of candidate models and next, one uses these $J_p + J_o$ models as input for the approach described above. Sensitivity to the choice of the particular DR estimator might be included by using a number of different DR estimators and selecting among or averaging over DR estimators that give similar estimates $\hat{\mu}_{i^*j^*}$.

van der Laan (2005) has proposed some new approaches to model selection for DR estimation that go beyond his above-mentioned approach, which we do not discuss here due to space limitations. Finally, Wang, Petersen, Bangsberg and van der Laan (2006) have proposed using the parametric bootstrap to study the sensitivity of DR estimates to highly variable "inverse probability" weights.

## 7. FURTHER CONSIDERATIONS

### Estimation of Causal Effects

K&S briefly touch on the problem of estimating the difference of the outcome means corresponding to two treatments in an observational study under ignorability. This difference is often referred to as the average causal effect (ACE). K&S view the problem of estimating ACE essentially as two missing-data problems, each one regarding the outcomes of subjects that do not follow the treatment of concern as missing. The difference of the DR estimators of the separate means serves as an estimator of the mean difference ACE. However, the difference of the two DR estimators will have poor small sample behavior if there is incomplete overlap of the estimated propensity scores in the treated and untreated. In fact, in the presence of incomplete overlap, most investigators argue against trying to estimate ACE and in favor of estimating the causal effect in the subpopulation of subjects with overlapping propensity scores. However, assuming the ACE parameter is of some substantive interest, Robins et al. (2007) suggest an alternative to reporting the difference of two DR estimators of the separate means. Their approach is based on fitting a linear semiparametric regression model for the unknown conditional effect function encoding the dependence of the conditional treatment effect on the baseline covariates $X$. Their model has the property that it is guaranteed to be correctly specified under the null hypothesis that the conditional effect function is the zero function. Robins et al. (2007) show that this strategy results in estimators of the ACE that greatly outperform any estimator based on the difference of double-robust estimators, whenever the model for the conditional effect function is correctly specified; in particular, when the aforementioned null hypothesis is true.

### Multiple Robustness

Consider again the MAR missing-data model with $X$ very high dimensional (say 20–100 continuous covariates) so we cannot possibly hope to model the propensity score or the outcome regression nonparametrically. Double-robust estimators of the mean $\mu$ of $Y$ are $n^{1/2}$-consistent if either one of two parametric models is correct but inconsistent if both models are misspecified. This property of DR estimators seems unsatisfactory, as it means that one does very, very well if one of the two models is correct but can do very, very poorly when both are incorrect. Might we do better?

Define an estimator to be *m-robust* for $\mu$ at rate $n^\alpha$ if the estimator is $n^\alpha$-consistent for $\mu$ when any one of $m$ parametric models is correct, but inconsistent if all $m$ models are misspecified. A DR estimator is then a 2-robust estimator with $\alpha = 1/2$. Our view is that an $m$-robust estimator with $m$ large, even though this may require $\alpha$ to be much smaller than $1/2$ and so entail a much slower rate of convergence, would usually be preferable to a DR estimator for the following two reasons. First, if one uses an $m$-robust rather than a DR estimator, one is more likely to be using a consistent estimator of $\mu$ (as it is always more likely that at least one of $m$, rather than one of two, models is correct). Second, the slower rate of convergence (under the assumption one of the $m$ models is correct) will result in wider nominal confidence intervals than the usual nominal intervals of length $1/n^{1/2}$ associated with a DR estimator. Such a wide interval seems to us a more appropriate measure of the actual uncertainty about $\mu$, more accurately reflecting the fact that our estimator could even be inconsistent if all $m$ models are incorrect.

These observations raise the following questions. Do *m-robust* estimators exist for arbitrarily large $m$ if we are willing to sacrifice $n^{1/2}$-*consistency* for $n^\alpha$-consistency with $\alpha$ perhaps much smaller than $1/2$? What is the maximum value of $\alpha$ we can achieve for a given $m$? If *m-robust* estimators exist for $m > 2$, how do we construct them? Answers to these questions can be found in Robins, Li, Tchetgen and van der Vaart (2007), where it is shown that, under weak assumptions, (i) *m-robust* estimators exist for all $m$, (ii) $m$-robust estimators are $(m-1)$ dimensional U-statistics, for which explicit closed-form expressions are given, and (iii) the maximal possible $\alpha$ is often less than $1/2$ and sometimes much less. However, the finite sample properties of *m-robust* estimators have yet to be studied even by simulation. Thus we will have to wait to see if they are as useful in practice as theory would indicate they should be.

# REFERENCES

Bang, H. and Robins, J. M. (2005). Doubly robust estimation in missing data and causal inference models. *Biometrics* **61** 962–972. MR2216189

Robins, J. M. (2000). Robust estimation in .sequentially ignorable missing data and casual inference models. *Proc. Amer. Statist. Assoc. Section on Bayesian Statistical Science* **1999** 6–10.

Robins, J. M. (2002). Commentary on "Using inverse weighting and predictive inference to estimate the effects of time-varying treatments on the discrete-time hazard", by Dawson. and Lavori. *Statistics in Medicine* **21** 1663–1680.

Robins, J. M. and Rotnitzky, A. (2001). Comment on "Inference for semiparametric models: Some questions and an answer," by P. J. Bickel and J. Kwon. *Statist. Sinica* **11** 920–936. MR1867326

Robins, J. M. and Wang, N. (2000). Inference for imputation estimators. *Biometrika* **87** 113–124. MR1766832

Robins, J. M., Rotnitzky, A. and Zhao L.-P. (1995). Analysis of semiparametric regression models for repeated outcomes in the presence of missing data. *J. Amer. Statist. Assoc.* **90** 106–121. MR1325118

Robins, J. M., Li, L., Tchetgen, F. and van der Vaart, A. W. (2007). Higher order influence functions and minimax estimation of nonlinear functionals. In *IMS Lecture Notes–Monograph Series Probability and Statistics Models*: *Essays in Honor of David A. Freedman* **2** 335–421.

Robins, J. M., Sued, M., Lei-Gomez, Q. and Rotnitzky, A. (2007). Double-robustness with improved efficiency in missing and causal inference models. Technical report, Harvard School of Public Health.

Rosenbaum, P. R. (2002). *Observational Studies*, 2nd ed. Springer, New York. MR1899138

Scharfstein, D. O., Rotnitzky, A. and Robins, J. M. (1999). Adjusting for nonignorable drop-out using semiparametric non-response models (with discussion). *J. Amer. Statist. Assoc.* **94** 1096–1146. MR1731478

Tan, Z. (2006). A distributional approach for causal inference using propensity scores. *J. Amer. Statist. Assoc.* **101** 1619–1637. MR2279484

van der Laan, M. J. and Robins, J. M. (2003). *Unified Methods for Censored Longitudinal Data and Causality*. Springer, New York. MR1958123

van der Laan, M. J. and Rubin, D. (2006). Targeted maximum likelihood learning. U.C. Berkeley Division of Biostatistics Working Paper Series. Available at http://www.bepress.com/ucbbiostat/paper213/.

van der Laan, M. (2005). Statistical inference for variable importance. U.C. Berkeley Division of Biostatistics Working Paper Series. Available at http://www.bepress.com/ucbbiostat/paper188/.

Wang, Y., Petersen, M., Bangsberg, D. and van der Laan, M. (2006). Diagnosing bias in the inverse probability of treatment weighted estimator resulting from violation of experimental treatment assignment. U.C. Berkeley Division of Biostatistics Working Paper Series. Available at http://www.bepress.com/ucbbiostat/paper211/.